\documentclass{jfm}
\usepackage{bm}
\usepackage{graphicx}
\usepackage{amsmath}
\usepackage{color}
\usepackage{multicol}
\usepackage{epstopdf}
\usepackage{subfig}

\newcommand\Wi{\mathit{Wi}}

\newcommand\Tr[1]{\operatorname{Tr}\left[#1\right]}
\allowdisplaybreaks
\newcommand{\rem}[1]{}

\begin{document}

\shorttitle{Lyapunov dimension of elastic turbulence} %for header on odd pages
\shortauthor{E. L. C. VI M. Plan, A. Gupta, D. Vincenzi and J. D. Gibbon} %for header on even pages
\title{Lyapunov dimension of elastic turbulence}
\author
 {Emmanuel Lance Christopher VI Medillo Plan\aff{1},
Anupam Gupta\aff{2},
Dario Vincenzi\aff{1}
\and
John D. Gibbon\aff{3}
%   \corresp{\email{xxx@xxx.xx}} 
}
\affiliation
{\aff{1}Universit\'e C\^ote d'Azur, CNRS, LJAD, Nice, France
%Universit\'e Nice Sophia Antipolis, Laboratoire J. A. Dieudonn\'e, 06108 Nice,
\aff{2} FERMaT, Universit\'e de Toulouse, CNRS, INPT, INSA, UPS, Toulouse, France
\aff{3} Department of Mathematics, Imperial College London, London SW7 2AZ, UK
}
\maketitle

\begin{abstract}
Low-Reynolds-number polymer solutions exhibit a chaotic behaviour known as `elastic turbulence' when the Weissenberg number exceeds a critical value. The two-dimensional Oldroyd-B model is the simplest constitutive model that reproduces this phenomenon. To make a practical estimate of the resolution scale of the dynamics requires an assumption that an attractor of the Oldroyd-B model exists\,: numerical simulations show that the quantities on which this assumption is based are bounded. We estimate the Lyapunov dimension of this assumed attractor as a function of the Weissenberg number by combining a mathematical analysis of the model with direct numerical simulations.
\end{abstract}
 
%%%%%%%%%%%%%%%%%%%%%%%
\section{Introduction}\label{sec:intro}

One of the most remarkable properties of viscoelastic fluids is the formation of instabilities at very low Reynolds numbers \citep{L92,S96}.
Such instabilities are of a purely elastic nature\,; they occur when inertial forces are negligible and elastic forces are strong. In polymer 
solutions, elastic instabilities eventually lead to a chaotic regime known as elastic turbulence \citep{GS00}.  The emergence of this regime is characterized by a fast growth of the Lyapunov exponent of the flow as polymer elasticity exceeds a critical value \citep{BSS04}.  In addition, 
the kinetic-energy spatial and temporal spectra have a power-law behaviour, which indicates the presence of a large number of active scales 
in the flow. The spatial spectrum, however, is steeper than for Newtonian turbulence, i.e. velocity fluctuations are concentrated at small 
wave numbers \citep{GS00,BSS07}.

Elastic turbulence has important applications in microfluidics in view of the fact that it strongly enhances mixing in devices that, owing to 
their microscopic size, are characterized by a low Reynolds number \citep{GS01}. Moreover, the potential use of elastic turbulence in the 
oil industry has recently emerged as a promising application. Aqueous polymer solutions are indeed used to recover the oil that remains 
trapped inside the pores of reservoir rocks after an initial water flooding, and elastic turbulence has been proposed as a mechanism to explain 
the unexpectedly high efficiency of this oil recovery method~\citep{MLHC16}.

The simplest constitutive model of polymer solutions is the Oldroyd-B model \citep{O50} in which the dissolved polymer phase is described by a symmetric tensor field, termed the polymer conformation tensor, which represents the moment of inertia of polymers averaged 
over thermal fluctuations. The Oldroyd-B model is thus a system of partial differential equations (PDEs) that describes the joint evolution of the velocity and the polymer conformation tensor. In particular, the polymer feedback on the flow is given by a stress term proportional to the conformation tensor. The relevant dimensionless number is the Weissenberg number\,; i.e. the ratio of the polymer relaxation time and the typical time scale of the flow. The main limitation of this model is that it assumes linear polymer elasticity which, in extensional flows and in the absence of polymer feedback, can lead to an unbounded growth of the conformation tensor and hence of polymer stresses.

In spite of its simplicity, the model successfully reproduces the main features of elastic turbulence \citep{FL03,BBBCM08,BB10,GVQE13}. In particular, numerical simulations show that elastic turbulence is observed also in two-dimensional settings. It has attracted much attention in the mathematical community over the last two decades in which analysts have focused on the existence, uniqueness and regularity of solutions~\citep[see, e.g.,][for some recent studies]{LMZ10,BS2011,CK12,ER15,HL16}. To our knowledge there are few theoretical results on the properties of elastic turbulent solutions.

A mathematical definition of the number of degrees of freedom of a system
of PDEs is given by the dimension of its global attractor \citep{R01}.
For the two-dimensional case the bounds found by \cite{CK12} on the $L^2$-norms of the vorticity $\omega$ and the polymer conformation tensor $\bm\sigma$, $\|\omega\|_2$ and $\|\bm\sigma\|_2$, are exponential in time (and double exponential for $\|\bnabla\bm\sigma\|_{2}$). Thus, no bounded long-time averages $\left<\cdot\right>$ have been found to exist and therefore, 
in a strictly rigorous sense, no global attractor is known to exist. Mathematically we can go no further. However, numerical simulations of elastic turbulence (see \S\ref{sec:numerical}) suggest that $\|\omega\|_{2}$, $\|\bm\sigma\|_{2}$ and $\|\bnabla\bm\sigma\|_{2}$ are indeed bounded in time for various values of $\Wi$.  One practical way of progressing is to work under the following technical assumption with a subsequent strategy\,:
\vspace{3mm}
\begin{enumerate}\itemsep 2mm
\item Given that numerical calculations of $\|\omega\|_{2}$, $\|\bm\sigma\|_{2}$ and $\|\bnabla\bm\sigma\|_{2}$ are finite in time, as suggested in \S\ref{sec:numerical}, we assume that a global attractor $\mathcal{A}$ exists\,;

\item Based on (i), we estimate the Lyapunov dimension of $\mathcal{A}$, which will use the long-time averages $\left<\|\omega\|_{2}\right>$, 
$\left<\|\bm\sigma\|_{2}\right>$ and $\left<\|\bnabla\bm\sigma\|_{2}\right>$\,: the numerical bounds found in \S\ref{sec:numerical} for these quantities  in terms of $\Wi$ are used in our estimates. 
\end{enumerate}
\vspace{3mm}
A connection between the system dynamics and the attractor dimension is provided by the notion of the Lyapunov exponents via the Kaplan--Yorke
formula.  For ordinary differential equations (ODEs), the Lyapunov exponents control the exponential growth or contraction of volume elements in phase space, and the Kaplan--Yorke formula expresses the balance between volume growth and contraction realized on the attractor.   The Kaplan--Yorke formula is used to define the Lyapunov dimension of the attractor and is the following\,: for Lyapunov exponents $\mu_1\geq\mu_2\geq\cdots\geq\mu_{n}\geq\cdots$, the Lyapunov dimension 
$d_{L}(\mathcal{A})$ is given by
\begin{equation}\label{dldef}
d_{L}(\mathcal{A}) = N -1 + \frac{\mu_{1}+\ldots + \mu_{N-1}}{-\mu_{N}},
\end{equation}            
where the number $N$ of $\mu_{n}$ is chosen to satisfy
\begin{equation}\label{KY2}
\sum_{n=1}^{N-1}\mu_{n}\geq 0
\hspace{1cm}\mbox{but}\hspace{1cm}
\sum_{n=1}^{N}\mu_{n} < 0\,.
\end{equation}
Note that according to the definition of $N$, the ratio of exponents in \eqref{dldef} satisfies
\begin{equation}\label{KY1}
0 \leq \frac{\mu_{1}+\ldots + \mu_{N-1}}{-\mu_{N}} < 1\,.
\end{equation}          
In simple terms, the value of $N$ that turns the sign of the sum of the Lyapunov exponents as in \eqref{KY2} is that value that bounds above $d_{L}(\mathcal{A})$ such that
\begin{equation}\label{KY3}
N-1 \leq d_{L}(\mathcal{A}) < N\,. 
\end{equation}
The non-integer Lyapunov dimension $d_{L}(\mathcal{A})$ generally bounds above the fractal and Hausdorff dimensions $d_{F}(\mathcal{A})$ and $d_{H}(\mathcal{A})$. 
The Kaplan--Yorke formula originated as a phase-space argument for ODEs but has been rigorously applied to global attractors in PDEs by \citet{CF85}~\citep[see also ][]{GT97}. To use the method for PDEs to find an estimate for $d_{L}(\mathcal{A})$, it is necessary to extend the idea of the Lyapunov exponents to {\em global} Lyapunov exponents via a trace formula for the two-dimensional Oldroyd-B model that is explained in \S\ref{sec:fractal}. This uses the methods of \cite{CF85}, \cite{CFT88}, \cite{DG91} and, in particular, the $L^{\infty}$-estimates of \cite{Const87}. For a two-dimensional system of side 
$L$, the resolution length $\ell_{\mathrm{res}}$ of the smallest feature in the dynamics is connected to $d_L(\mathcal{A})$ by $d_L(\mathcal{A}) \sim \left(L/\ell_{\mathrm{res}}\right)^{2}$.

%%%%%%%%%%%%%%%%%%%%%%%%%%%%%%%
\section{The Oldroyd-B model for polymer solutions}\label{sec:model}

On the periodic unit square $\Omega = [0,\,1]^{2}$, the dimensionless form of the Oldroyd-B model is
\begin{subeqnarray}
\partial_t {\bm u} + {\bm u}\bcdot\bnabla {\bm u} &=& -\bnabla p + \Rey^{-1} \Delta{\bm u} + 
\beta(\Wi\,\Rey)^{-1} \bnabla \bcdot {\bm\sigma} + {\bm F}\,,\slabel{eq:NSE1}\\
\partial_t{\bm\sigma} + {\bm u} \bcdot \bnabla {\bm\sigma} &=& (\bnabla {\bm u}){\bm\sigma}
 + {\bm\sigma}(\bnabla {\bm u})^\top
 - \Wi^{-1}({\bm\sigma}- \mathsfbi I) + \Pen^{-1} \Delta{\bm\sigma},\slabel{eq:stress1}
\label{eq:sys1}
\end{subeqnarray}
where $\bm u$ is the incompressible velocity field, $(\bnabla u)_{ij}=\partial_j u_i$, $\bm\sigma$ is the polymer conformation tensor, $p$ is pressure, and where 
$\Rey$, $\Pen$ and  $\Wi$ are the Reynolds, P\'eclet and Weissenberg numbers respectively. The positive constant $\beta$ 
depends on the polymer concentration and equilibrium extension. The forcing $\bm F$ is time-independent, periodic and 
divergence-free ($\bnabla \bcdot \bm F=0$).

The Laplacian term in \eqref{eq:stress1} originates from the diffusion of the centre of mass of polymers \citep{EL89}, which ensures 
the global regularity of the two-dimensional Oldroyd-B model \citep{CK12} and improves the stability of numerical simulations, even though the values of $\Pen$ used in practice are considerably lower than its realistic values \citep{SB95,T11}.

In two dimensions, \eqref{eq:sys1} can be rewritten in terms of the scalar vorticity $\omega=\hat{\bm z}\bcdot(\bnabla\times\bm u)$\,:
\begin{subeqnarray}
\partial_t \omega + {\bm u}\bcdot \bnabla \omega &=& \Rey^{-1} \Delta \omega + \beta(\Wi\,\Rey)^{-1}\,
\hat{\bm z}\bcdot \bnabla \times (\bnabla \bcdot {\bm\sigma})+ f,
\slabel{eq:NSE2}
\\
\partial_{t}{\bm\sigma} + {\bm u} \bcdot \bnabla {\bm\sigma} 
&=&
(\bnabla {\bm u}){\bm\sigma}
 + {\bm\sigma}(\bnabla {\bm u})^\top
 - \Wi^{-1}({\bm\sigma}-\mathsfbi{I}) + \Pen^{-1}\Delta{\bm\sigma}\,,
\slabel{eq:stress2}
\label{eq:sys2}
\end{subeqnarray}
where $f=\hat{\bm z}\bcdot (\bnabla \times {\bm F})$ and $\bm u=\bnabla^\perp\Delta^{-1}\omega$ with $\bnabla^\perp\equiv(-\partial_y,\partial_x)$.
Since $\bm F$ is periodic, the spatial average of $\omega$ is zero, and the inverse Laplacian of $\omega$ is properly defined. Figure 1 shows snapshots of the $\omega$ and $\operatorname{Tr}\bm\sigma$ fields from a numerical simulation of \eqref{eq:sys2} in the elastic-turbulence regime. 
\begin{figure}
  \centerline{(a)\includegraphics[width=0.46\textwidth]{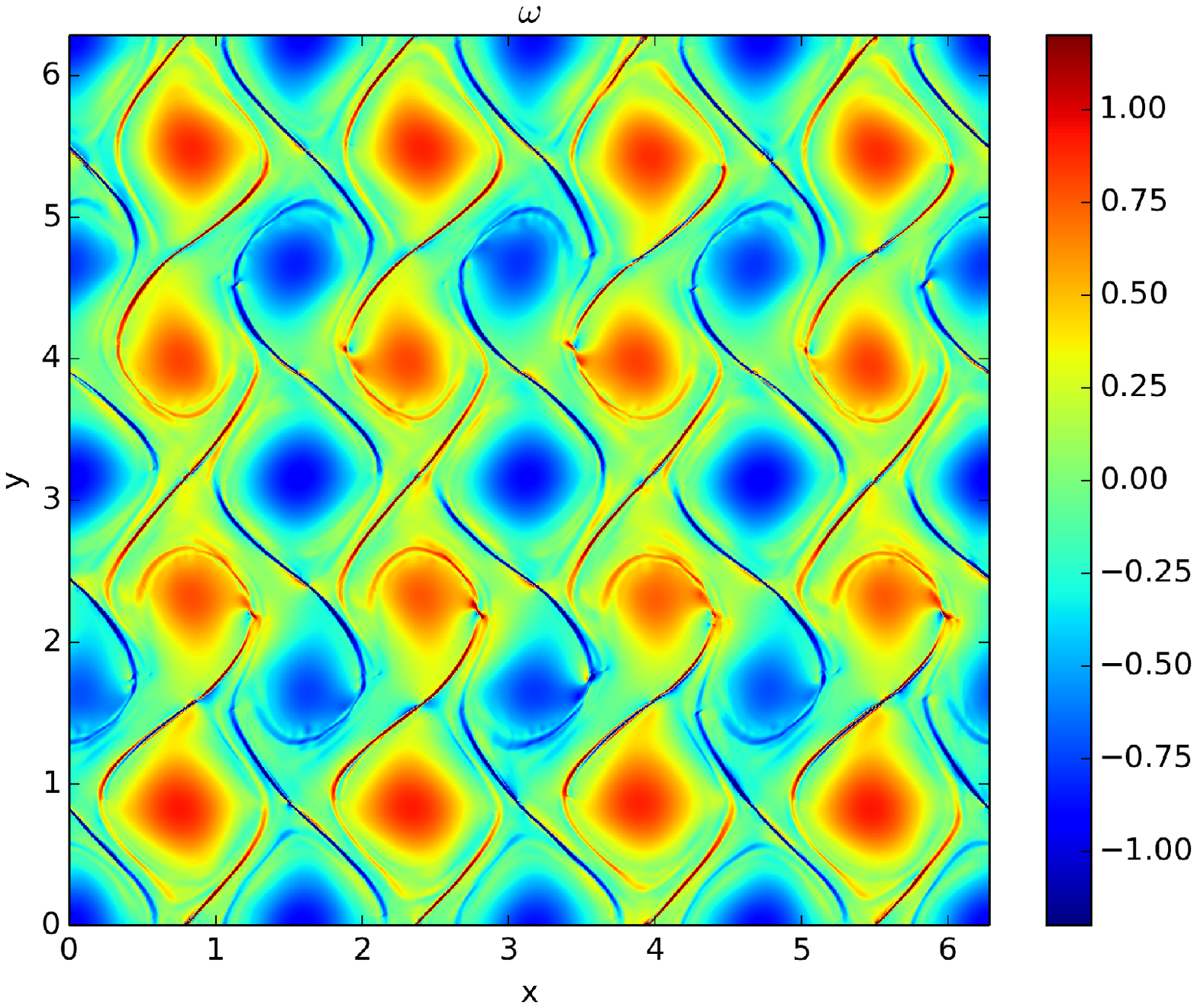} 
  (b)\includegraphics[width=0.46\textwidth]{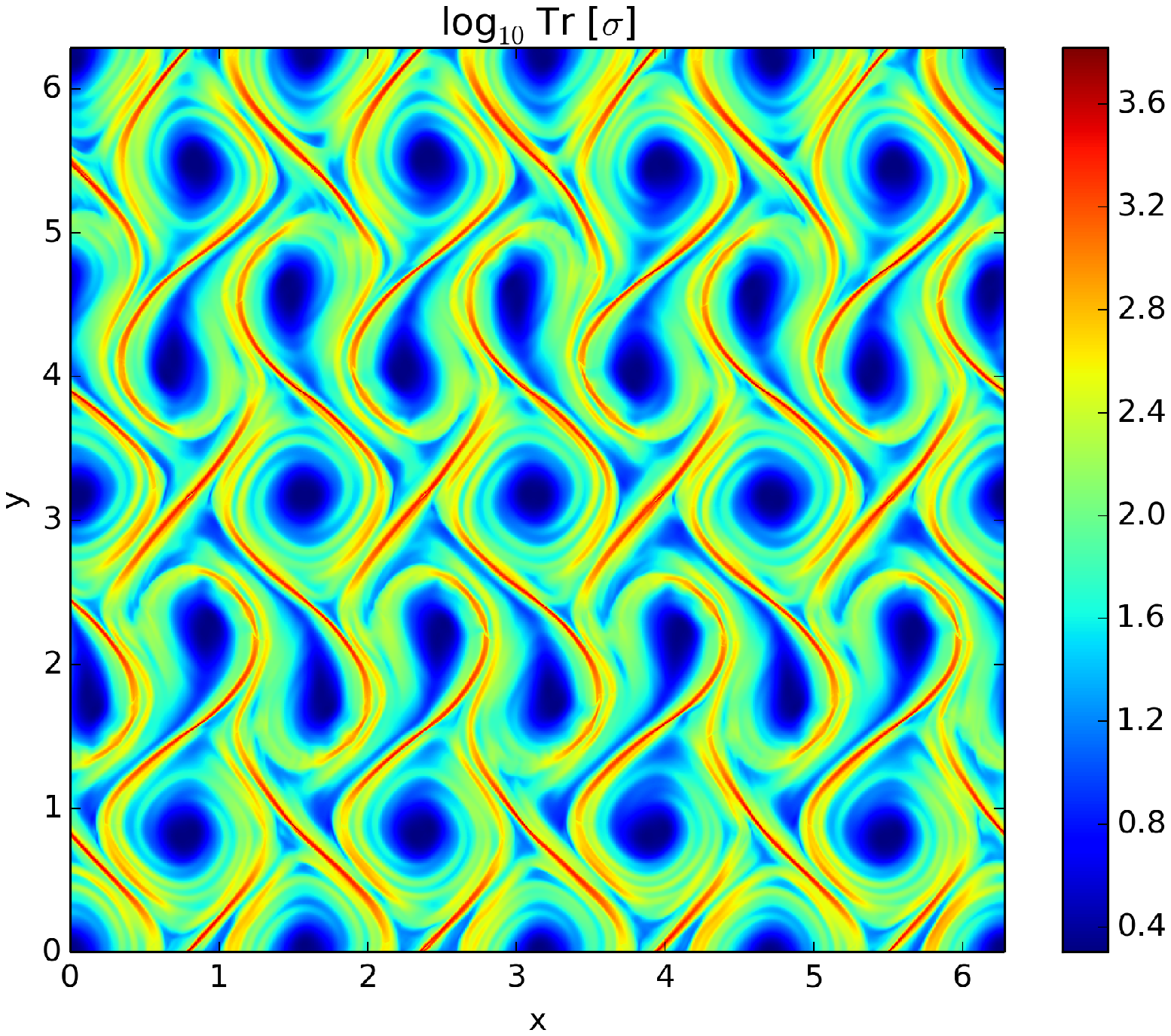}}
  \caption{Pseudocolor plot of\,: (a) $\omega$ and (b) $\log_{10}\operatorname{Tr}\bm\sigma$ for $\Rey=\Rey_\mathrm{c}/\sqrt{2}$, $\Wi=20$, $\beta=0.2$\, and a cellular forcing with $n=4$ and $f_0=0.16$ (see \S\ref{sec:numerical} for the details).}
\label{fig:snapshot}
\end{figure}
Note that $\operatorname{Tr}\bm\sigma$ is concentrated over very thin regions, which is associated with large gradients in the polymer-conformation-tensor field.

%%%%%%%%%%%%%%%%%%%%%%%%%%%
\section{Lyapunov dimension}\label{sec:fractal}

In the PDE case the phase space is now infinite-dimensional. Define ${\bm q} = (\omega,\bm\sigma)$ and denote by $\delta {\bm q} = (\delta\omega,\delta\bm\sigma)$ the  infinitesimal displacement about ${\bm q}$\,; $\delta{\bm q}$ satisfies the linearized set of equations from \eqref{eq:sys2}:
\begin{gather}
\delta{\dot{\bm q}}
=  {\mathsfbi{A}}(t)\delta{\bm q}\\
=
\begin{pmatrix}
-{\bm u}\bcdot \bnabla\delta\omega
+\dfrac{1}{\Rey} \Delta\delta\omega 
- \delta{\bm u} \bcdot \bnabla \omega 
+\dfrac{\beta}{\Wi\,\Rey}{\hat{\bm z}}\bcdot \bnabla \times (\bnabla\bcdot\delta\bm\sigma) \vspace{0mm}\\
(\bnabla\delta{\bm u}){\bm\sigma}
+{\bm\sigma}(\bnabla\delta{\bm u})^\top
-\delta{\bm u}\bcdot\bnabla{\bm\sigma}
-{\bm u}\bcdot\bnabla\delta\bm\sigma
+(\bnabla{\bm u})\delta\bm\sigma
+\delta\bm\sigma(\bnabla {\bm u})^\top
-\dfrac{\delta\bm\sigma}{\Wi}
+\dfrac{1}{\Pen}\Delta\delta\bm\sigma
\end{pmatrix},
\label{eq:terms}\nonumber
\end{gather}
where the explicit form of the operator $\mathsfbi A(t)$ is obtained by using $\bm u=\bnabla^\perp\Delta^{-1}\omega$ to express 
$\bm u$ and $\delta\bm u$ in terms of $\omega$ and $\delta\omega$, respectively. Following \cite{CF85}, take different sets of 
initial conditions ${\bm q}(0)+\delta{\bm q}_{i}(0)$ which evolve into ${\bm q}(t)+\delta{\bm q}_{i}(t)$ for $i =1,\ldots,N$.  If they 
are chosen to be linearly independent, these $\delta{\bm q}_{i}$ form an $N$-volume or parallelepiped of volume $V_{N}(t) = \left|\delta{\bm q}_{1}\wedge\delta{\bm q}_{2} \ldots\wedge\delta{\bm q}_{N}\right|\,,$
which changes along the solution $\bm q(t)$, so we need to find its time evolution.  
This is given by $\dot{V}_{N} = V_{N}\operatorname{Tr}\left[\mathsfbi{A}\mathsfbi{P}_{N}\right]$, 
which is easily solved to give  
\begin{equation}\label{trace2} 
V_{N}(t) = V_{N}(0) \exp\left\{\int_{0}^{t}\operatorname{Tr}\left[\mathsfbi{A}\mathsfbi{P}_{N}\right](\tau)\,\mathrm{d}\tau\right\}\,,
\end{equation}
where $\mathsfbi{P}_{N}(t)$ is an $L^{2}$-orthogonal projection onto the finite-dimensional subspace
$\mbox{span}\left\{\delta{\bm q}_{1},\delta{\bm q}_{2},\ldots,\delta{\bm q}_{N}\right\}$.  
In terms of the time average $\left<\cdot\right>$, the sum of the first $N$ global Lyapunov exponents is taken to be \citep{CF85}
\begin{equation}\label{global1}
\sum_{n=1}^{N}\mu_{n} = \left<\operatorname{Tr} \left[\mathsfbi{A}\mathsfbi{P}_{N}\right]\right> .
\end{equation}
To estimate the Lyapunov dimension, we wish to find the value of $N$ that turns the sign of $\left<\operatorname{Tr}\left[\mathsfbi{A}\mathsfbi{P}_{N}\right]\right>$. This value of $N$ bounds above $d_{L}(\mathcal{A})$ as in \eqref{KY3}. 
As we are interested in elastic turbulence,
we assume $\Pen\gg1$, $\Wi\gg1$ and $0<\Rey<\Rey_{\mathrm{c}}$, where $\Rey_{\mathrm{c}}$ is the critical value for the appearance of inertial instabilities. The derivation has been relegated to the Appendix. The final result is that the following
inequalities are sufficient conditions on $N$ for the growth rate
$\left<\operatorname{Tr}\left[\mathsfbi{A}\mathsfbi{P}_{N}\right]\right>$
to be negative and the $N$-volume $V_N$ to contract\,: 
\begin{subeqnarray}
N&>& c\Rey\langle\|\bm\sigma\|_2^3\rangle^{1/3}
\big(1+\ln \Rey+\ln\langle\|\bm\sigma\|_2^3\rangle^{1/3}\big)^{1/2},
\slabel{eq:estsigma}\\
%\mathcal{N} &\geq& c\Rey\langle\|\omega\|_2^2\rangle^{1/2}
%\slabel{eq:estomega}\\
%\mathcal{N} &\geq& c\Rey^{4/3}\|f\|_2^{2/3}
%\left(1+\ln \Rey^2+\ln\|f\|_2\right)^{1/3}\\
N&>& 
%\left(c+\dfrac{\beta}{\Wi}\right)^{2/3}
c\Rey^{2/3}\langle\|\bnabla\bm\sigma\|_2^2\rangle^{1/3}
\big(1+\ln \Rey+\ln\langle\|\bnabla\bm\sigma\|_2^2\rangle^{1/2}\big)^{1/3}\,,
\slabel{eq:estgradsigma}
\label{eq:estimation}
\end{subeqnarray}
where $\|\cdot\|_{2}^{2} = \int_\Omega |\cdot|^{2}\mathrm{d}\bm x$, $|\bm\sigma|^{2} = \sum_{i,j}|\sigma_{ij}|^{2}$, $|\bnabla\bm\sigma|^2=\sum_{i,j,k}|\partial_k\sigma_{ij}|^{2}$ and $c$ is regarded as a generic, positive, dimensionless constant.
These conditions depend on estimates for $\left<\|\bm\sigma\|_{2}^3\right>$ and $\left<\|\bnabla\bm\sigma\|_{2}^2\right>$.  It has been shown in \cite{CK12} that these are exponential in time and thus cannot be used. Instead we turn to direct numerical simulations to find their behaviour in terms of $\Wi$. 

%%%%%%%%%%%%%%%%%%%%%
\section{Direct numerical simulations of elastic turbulence}\label{sec:numerical}

To simulate solutions of \eqref{eq:sys2}, we use the approach described in \citet{GPP15} and \citet{GP16}. For the time integration, we use 
the fourth-order Runge--Kutta scheme with timestep $\delta t=10^{-4}$, while the fourth-order central-finite-difference scheme with $1024^2$ collocation points is employed for the spatial derivatives. To accurately resolve the steep gradients of $\bm\sigma$, we apply the Kurganov--Tadmor scheme to the advection term in \eqref{eq:stress2} \citep{KT00} \citep[see][for the application of this scheme to viscoelastic fluids]{vaithi}, which allows us to set $\Pen=\infty$ in \eqref{eq:sys2}. The velocity is calculated from the vorticity via the Poisson equation $\Delta\psi=\omega$, where $\psi$ is the stream function: $\bm u=\bnabla^\perp\psi$. The pseudospectral method is used to solve the Poisson equation in Fourier space. The simulations were performed on $[0,2\upi]^2$ and the solutions were rescaled appropriately.

Three kinds of forcing are considered: a cellular forcing $f = -f_0 n[\cos(nx)+\cos(ny)]$ with $n=4$, a cellular forcing with $n=10$, and a Kolmogorov forcing $f=-f_0 n\cos(nx)$ with $n=8$. In all simulations $\Rey=\Rey_\mathrm{c}/\sqrt{2}$, where $\Rey_\mathrm{c}$ is the critical Reynolds number above which inertial instabilities occur; hence in the absence of polymers 
($\beta=0$) the stationary vorticity field shows alternating vortices and antivortices for the cellular forcing and a parallel sinusoidal flow for the
Kolmogorov forcing. The parameter $\beta$ is set to $\beta=0.2$.
We checked that in the elastic-turbulence regime the kinetic-energy transfer due to the fluid inertia is negligible, so the chaotic dynamics is entirely due
to polymer stresses.

\begin{figure}
\setlength{\unitlength}{\textwidth}
  \centerline{(a)\includegraphics[width=0.4\textwidth]{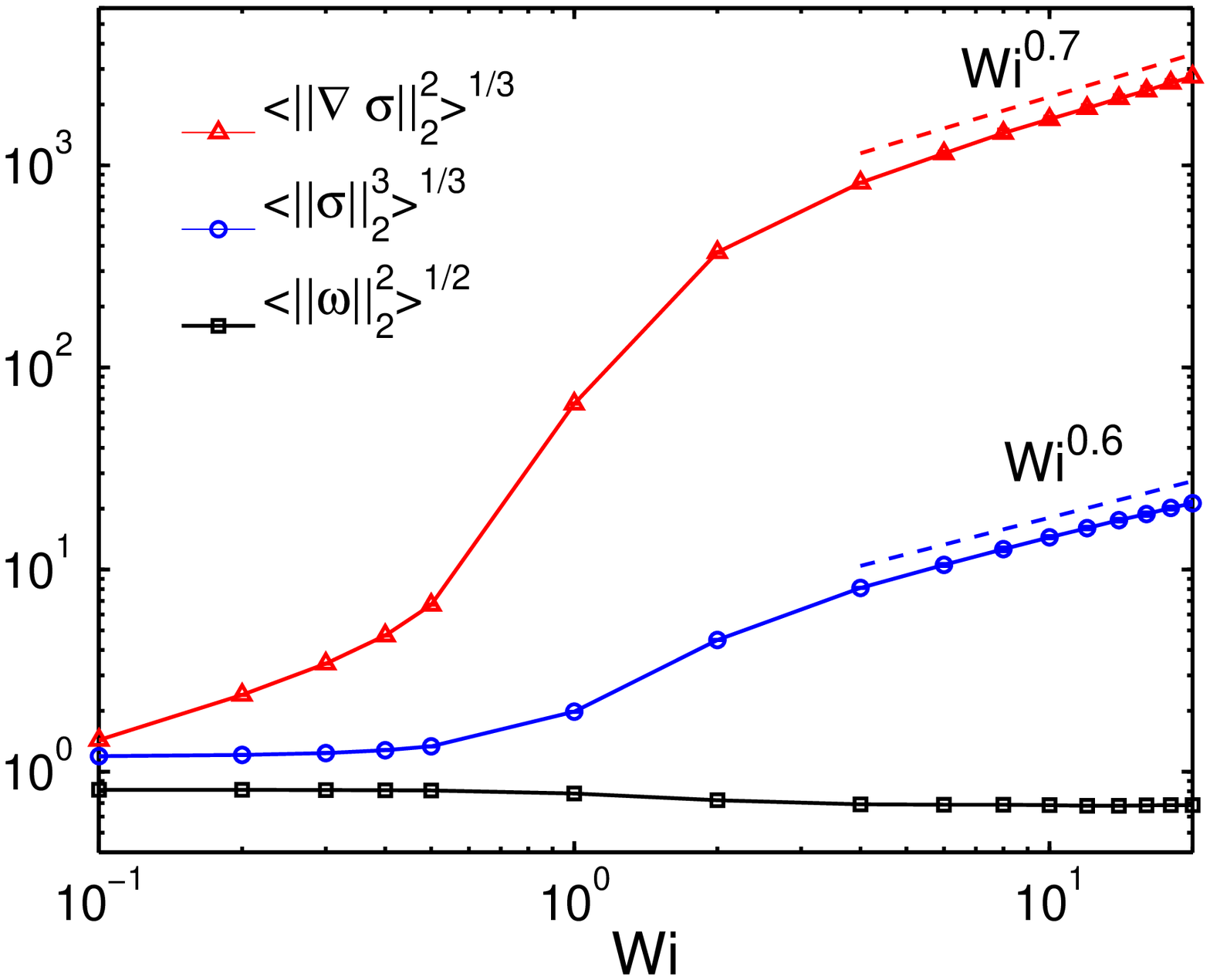}
  (b)\includegraphics[width=0.4\textwidth]{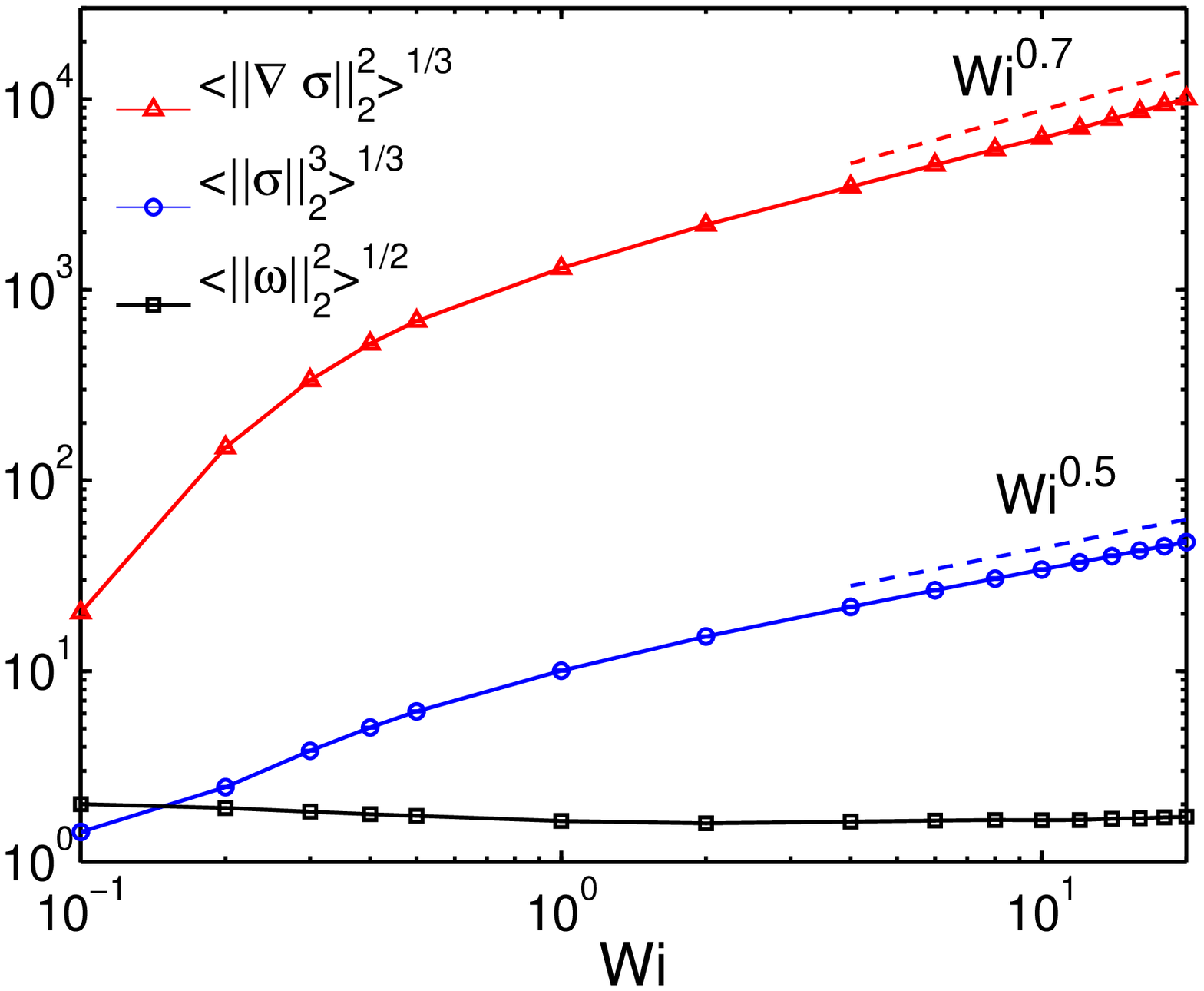}}
  \centerline{(c)\includegraphics[width=0.4\textwidth]{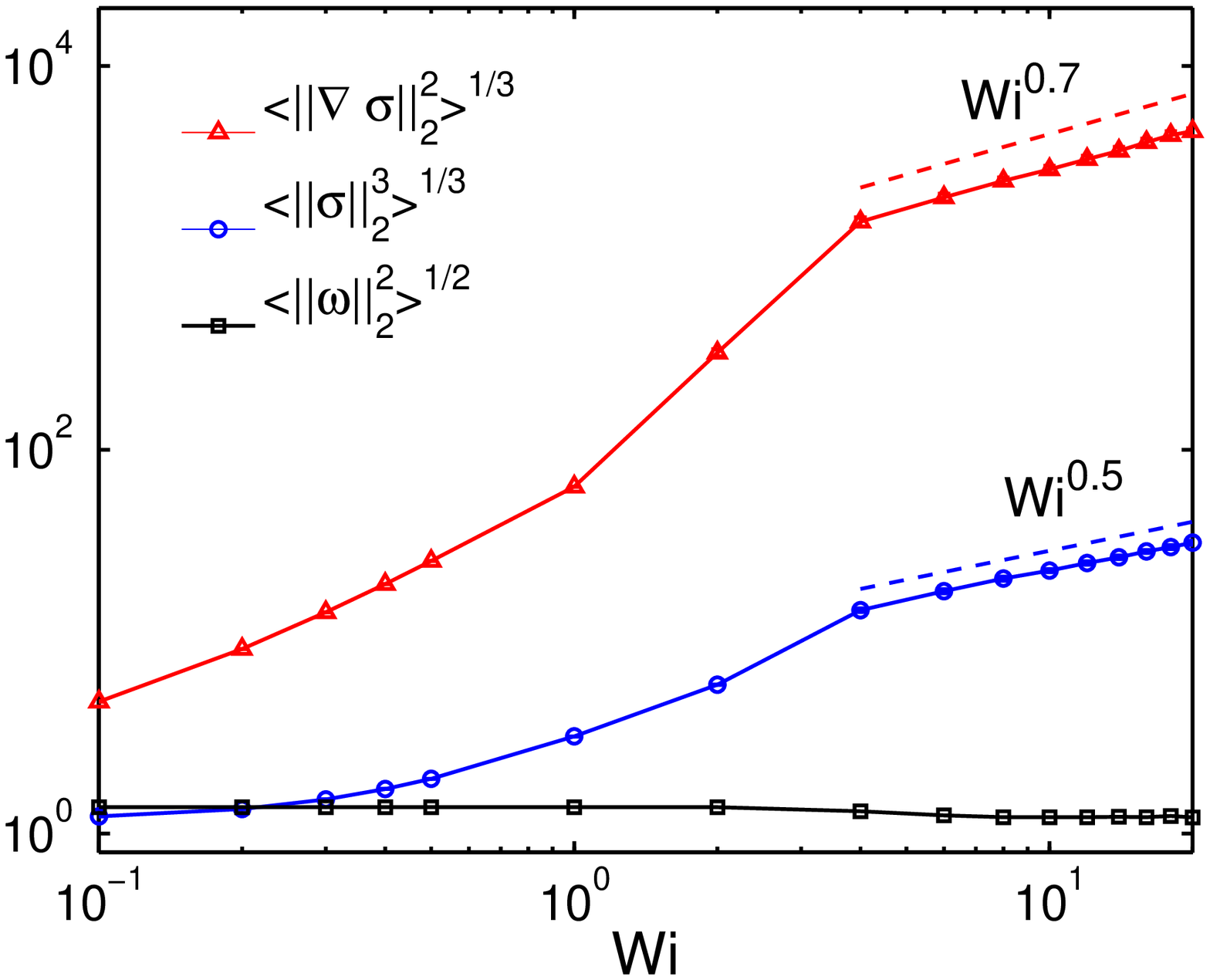}
  (d)\includegraphics[width=0.4\textwidth]{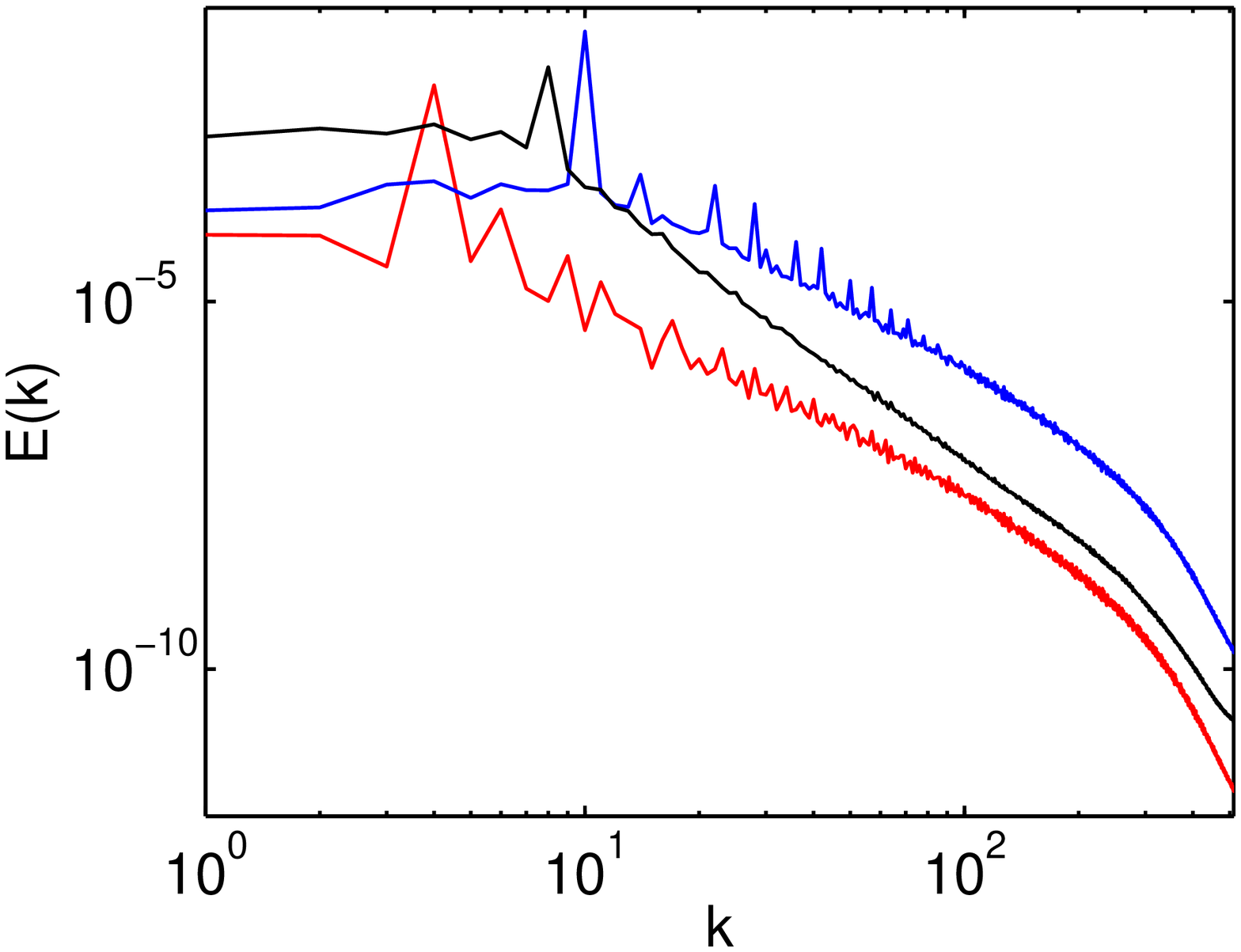}}
  \caption{Plot of $\langle\|\omega\|_2^2\rangle^{1/2}$ (black squares), $\langle\|\bm\sigma\|_2^3\rangle^{1/3}$ (blue circles) and 
$\langle\|\bnabla\bm\sigma\|_2^2\rangle^{1/3}$ (red triangles) as a function of $\Wi$ for $\Rey=\Rey_\mathrm{c}/\sqrt{2}$,  $\beta=0.2$, and (a) cellular forcing 
with $n=4$, $f_0=0.16$, (b) cellular forcing with $n=10$, $f_0=2.50$, (c) Kolmogorov forcing with $n=8$, $f_0=1.28$; (d) Kinetic-energy spectrum for $\Wi=20$ for case (a) in red, case (b) in blue and case (c) in black. The spectrum shows a power-law behaviour $E(k)\simeq k^{-a}$ with $a\approx3.1$ for case (a), $a\approx3.0$ for case (b) and $a\approx3.7$ for case (c).}
\label{fig:comparison}
\end{figure}

The numerical simulations show that 
$\langle\Vert\nabla\bm\sigma\Vert_2^2\rangle^{1/3}$ 
is greater than $\langle\Vert\bm\sigma\Vert_2^3\rangle^{1/3}$ 
and, at large $\Wi$, $\langle\Vert\nabla\bm\sigma\Vert_2^2\rangle^{1/3}\sim \Wi^\alpha$ with $\alpha\approx0.7$ for the three different forcings considered here (see figure \ref{fig:comparison}). 
This result is in agreement with the observation of large gradients in the polymer conformation field in figure~\ref{fig:snapshot} as well as in previous numerical simulations 
\citep{BBBCM08,BB10}.  It follows that the value of $N$ such that $N$-volumes contract is determined by inequality \eqref{eq:estgradsigma}. 
We conclude that, in the elastic-turbulence regime and under the specified assumptions, the two-dimensional Oldroyd-B model has a finite-dimensional global attractor with Lyapunov dimension
\begin{equation}
d_{L}(\mathcal{A}) \leq \mathcal{N}\,,
\label{finalestimate}
\end{equation}
where $\mathcal{N}$ is the minimum value of $N$ satisfying \eqref{eq:estgradsigma}.
Thus, up to logarithmic corrections,
\begin{equation}
d_{L}(\mathcal{A}) \lesssim c\Rey^{2/3}\langle\|\bnabla\bm\sigma\|_2^2\rangle^{1/3} \sim C\Wi^\alpha\,,
\label{eq:finalfinal}
\end{equation}
where $\alpha\approx0.7$ and $C$ depends on $\Rey$.
We have also performed numerical simulations at smaller values of
$\Rey$, which indicate that $\alpha$ does not depend on $\Rey$.
It is worth noting that since $\Vert\bm\nabla\bm\sigma\Vert$ grows
with $\Wi$, inequality \eqref{eq:estimation} implies that a lower $\Rey$ requires a higher $\Wi$ to obtain the same attractor dimension. This is in
agreement with the stability analysis of the Oldroyd-B model, according to which the critical $\Wi$ for the
appearance of elastic instabilities increases as $\Rey$ decreases \citep[e.g.][]{L92,BCMPV05}.

%%%%%%%%%%%%%
\section{Conclusions}
Through a mathematical and numerical analysis of the two-dimensional Oldroyd-B model in the elastic-turbulence regime, we have made a practical estimate of the dimension of its (assumed) global attractor, as suggested by numerical simulations.  We have also shown that the complexity of the attractor is related to the formation of large gradients in the polymer conformation field.
Although the asymptotic power-law dependence of $d_L(\mathcal{A})$ on $\Wi$ was found to be the same for different forcings, further studies are required to confirm the potential universality of the exponent $\alpha$.

As mentioned in the introduction, the main limitation of the Oldroyd-B model is the absence of a maximum polymer elongation. Other constitutive models of polymer solutions, such as the FENE-P model, overcome this limitation by introducing a non-linear elastic force that diverges when 
$\operatorname{Tr}\bm\sigma$ approaches the square of the maximum elongation \citep*[see][for the application of the FENE-P model to the 
study of elastic turbulence]{TST11,GP16}.  However, using a non-linear elastic force does not prevent the formation of large gradients in the $\bm\sigma$-field \citep*{TST11}. In other words, large values of $\bnabla\bm\sigma$ seem to occur independently
of the form of the force that describes the elasticity of polymers. 
Our estimate \eqref{eq:finalfinal} depends on $\|\bnabla\bm\sigma\|_2$ rather than $\|\bm\sigma\|_2$.
This fact suggests that, albeit based on the Oldroyd-B model, 
\eqref{eq:finalfinal} may also be relevant to other constitutive models.

Our analysis can be adapted to
the Oldroyd-B model coupled with the unsteady Stokes equations, in which the $\bm u\bm\cdot\bm\nabla\bm u$ term is set to zero;
the estimates for $N$ given in \eqref{eq:estimation} are unchanged. The $\Rey=0$ case, in which \eqref{eq:NSE1} is replaced
with the Stokes equations \citep{TS09,TST11},
would require by contrast a separate analysis, because $\omega$ would depend on $\bm\sigma$ via a time-independent differential relation and
\eqref{eq:sys2} would reduce to a dynamical system in the $\bm\sigma$-space only.
It would also be interesting to explore whether or not a relation 
exists between the exponent $\alpha$ in \eqref{eq:finalfinal} and the exponent of the singular structures that
emerge near to hyperbolic points in the $\Rey=0$ case \citep{R06,TS07}.

Elastic turbulence is observed at low $\Rey$ and high $\Wi$. When both $\Rey$ and $\Wi$ are high, the addition of polymers to a two-dimensional turbulent flow suppresses large-scale velocity fluctuations \citep{AK02}. This phenomenon is once again correctly reproduced by the Oldroyd-B model \citep{BCM03}. Our estimate \eqref{eq:estimateN_1}, which gives a sufficient condition for the contraction of $N$-volumes, holds for all values of $\Rey$. It would be interesting to investigate the implications of this estimate for the attractor dimension in the high-$\Rey$ regime.

\begin{acknowledgments}
The authors would like to acknowledge useful discussions with G. Boffetta, A. Mazzino, S. Musacchio, P. Perlekar and S.S. Ray. The work of ELCMP was supported by EACEA through the EMMA program. ELCMP and DV acknowledge the support of the EU COST Action MP 1305 `Flowing Matter.' The authors also acknowledge the Obervatoire de la C\^ote d'Azur for computing resources.
\end{acknowledgments}

\oneappendix
\section{Derivation of the decay rate of $N$-dimensional volumes}
\label{appA}
Here we show the main steps needed to obtain \eqref{eq:estimation}. The trace $\operatorname{Tr}\left[{\mathsfbi{A}}{\mathsfbi{P}}_N\right]$ 
can be expressed as \citep[e.g.][]{CF85,DG95,R01}
\begin{equation}
\operatorname{Tr}\left[{\mathsfbi{A}}{\mathsfbi{P}}_N\right]=\sum^N_{n=1} \int_\Omega \bm\Phi_n\bcdot\mathsfbi{A}\bm\Phi_n 
\, \mathrm{d}\bm x,
\label{eq:trace}
\end{equation}
where $\{\bm\Phi_n\}_{n=1}^N=\{(\phi_n^\omega,\bm\phi_n^\sigma)\}_{n=1}^N$ 
is an orthonormal set spanning the subspace generated by the displacements $\{(\delta\omega_n,\delta\bm\sigma_n)\}_{n=1}^N$
and where $\{\phi_n^\omega\}_{n=1}^{N}$ is such that $\int_\Omega\phi_n^\omega\phi_m^\omega\,\mathrm{d}\bm x=0$ if $n\neq m$ and 
 $\{\bm\phi_n^\sigma\}_{n=1}^{N}$ with $(\bm\phi_n^\sigma)^\top=\bm\phi_n^\sigma$ is such that $\int_\Omega \bm\phi_n^\sigma:\bm\phi_m^\sigma\,\mathrm{d}\bm x=0$ if $n\neq m$. 
The orthonormality of $\{\bm\Phi_n\}_{n=1}^N$ should then be interpreted as follows\,: $\int_\Omega \phi^\omega_m\phi^\omega_{n}\,
\mathrm{d}\bm x +\int_\Omega \bm\phi^\sigma_m :\bm\phi^\sigma_{n}\,\mathrm{d}\bm x=\delta_{mn},$ whence
\begin{equation}
\sum_{n=1}^N\int_\Omega|\bm\Phi_n|^2\mathrm{d}\bm x
=\sum_{n=1}^N\int_\Omega\Big(|\phi_n^\omega|^2+|\bm\phi_n^\sigma|^2\Big)\,\mathrm{d}\bm x = N\,.
\label{eq:orthonormalityofPhi}
\end{equation}
The symbol `:' denotes the inner product between matrices, $\bm\sigma:\bm\sigma'\equiv\operatorname{Tr}[\bm\sigma^\top\bm\sigma']$.
The following inequality \citep[e.g.][]{R01} will also be used when we estimate the norms of $\phi_n^\omega,\bm\phi_n^\sigma$ or of their gradients
\begin{equation}
\Theta \equiv \Tr{-\Delta \mathsfbi P_N} =
-\sum^N_{n=1}\int_\Omega \Big(\phi_n^\omega\Delta\phi_n^\omega + \bm\phi_n^{\sigma}:\Delta\bm\phi_n^{\sigma}\Big)\, \mathrm{d}\bm x \geq cN^2.
\label{eq:traceoflaplacian}
\end{equation}
The terms appearing in \eqref{eq:trace} can be explicitly listed as (the terms that follow from ${\bm u}\bcdot \bnabla\delta\omega$ and ${\bm u}\bcdot \bnabla\delta\bm\sigma$ vanish because of the incompressibility of $\bm u$)
\begin{subeqnarray}
\operatorname{Tr}\left[{\mathsfbi{A}}{\mathsfbi{P}}_N\right]
&=&
%%%%%%%%%%%%%%%%%%%%%%
\Rey^{-1}\sum^N_{n=1} \int_\Omega \phi_n^{\omega} \Delta \phi_n^\omega\,\mathrm{d}\bm x
+ \Pen^{-1}\sum^N_{n=1} \int_\Omega \bm\phi_n^{\sigma}:\Delta\bm\phi_n^{\sigma}\mathrm{d}\bm x 
\slabel{eq:trace1}\\
%%%%%%%%%%%%%%%%%%%%%%
&&\hspace{-1.6cm}
-\sum^N_{n=1} \int_\Omega \phi_n^{\omega} {\bm v}_n \bcdot \bnabla\omega \, \mathrm{d}\bm x
-\sum^N_{n=1} \int_\Omega \bm\phi_n^{\sigma}:({\bm v}_n \bcdot \bnabla {\bm\sigma}) \, \mathrm{d}\bm x 
\slabel{eq:trace2}\\
%%%%%%%%%%%%%%%%%%%%%%%%
&&\hspace{-1.6cm}
+ \beta(\Wi\,\Rey)^{-1} \sum^N_{n=1} \int_\Omega \phi_n^{\omega}{\hat{\bm z}}\bcdot(\bnabla \times \bnabla \bcdot \bm\phi_n^{\sigma}) \, \mathrm{d}\bm x
-\Wi^{-1}\sum^N_{n=1} \int_\Omega |\bm\phi_n^{\sigma}|^2 \,\mathrm{d}\bm x
\slabel{eq:trace3}\\
%%%%%%%%%%%%%%%%%%%%%%%%
&&\hspace{-1.6cm}
+\sum^N_{n=1} \int_\Omega
\bm\phi_n^{\sigma}:[{(\bnabla{\bm v}_n)\bm\sigma}+{\bm\sigma}(\bnabla{\bm v}_n)^{\top}]\,\mathrm{d}\bm x
+\sum^N_{n=1} \int_\Omega \bm\phi_n^{\sigma}:[(\bnabla{\bm u})\bm\phi_n^{\sigma}+\bm\phi_n^{\sigma}(\bnabla{\bm u})^\top] \,\mathrm{d}\bm x,\hspace{1cm}
\slabel{eq:trace4}
\label{eq:trace_1}
\end{subeqnarray}
where 
$\bm v_n=\bnabla^\perp\Delta^{-1}\phi_n^\omega$.
By using \eqref{eq:traceoflaplacian}, the Laplacian terms in \eqref{eq:trace1} can be shown to satisfy (under the assumption $Pe>\Rey$)
\begin{equation}
\Rey^{-1}\sum^N_{n=1} \int_\Omega \phi_n^{\omega} \Delta \phi_n^\omega\,\mathrm{d}\bm x
+ \Pen^{-1}\sum^N_{n=1} \int_\Omega\bm\phi_n^{\sigma}:\Delta\bm\phi_n^{\sigma}\,\mathrm{d}\bm x \leq -\Rey^{-1}\Theta\,.
\label{laplacianterms}
\end{equation}
We treat the advective terms \eqref{eq:trace2} by using the result of \citet{Const87} and its subsequent use in \cite{DG91}\,: i.e. a Cauchy--Schwarz inequality, an $L^\infty$ bound for the terms involving $\bm v_n$, and \eqref{eq:traceoflaplacian}
\begin{equation}
\left|\sum^N_{n=1} \int_\Omega \Big[\phi_n^{\omega} {\bm v}_n \bcdot \bnabla\omega + \bm\phi_n^{\sigma}:({\bm v}_n \bcdot \bnabla {\bm\sigma})\Big] \,\mathrm{d}\bm x \right| \leq c\,(1+\ln \Theta)^{1/2}\Theta^{1/4}(\|\bnabla\omega\|_2+\|\bnabla\bm\sigma\|_2)\,.
\label{advectiveterms}
\end{equation}
To estimate the feedback term in \eqref{eq:trace3}, we use integration by parts, the Cauchy--Schwarz inequality, the relations
$|\bnabla\times(\phi_n^{\omega}\hat{\bm z})| = |\bnabla\phi_n^{\omega}|$ and $|\bnabla\bcdot\bm\phi_n^{\sigma}|^2
\leq |\bnabla\bm\phi_n^{\sigma}|^2$, and \eqref{eq:traceoflaplacian}\,: 
\begin{equation}
\left|\sum^N_{n=1} \int_\Omega\phi_n^{\omega}{\hat{\bm z}}\bcdot(\bnabla\times\bnabla\bcdot\bm\phi_n^{\sigma})\,\mathrm{d}\bm x \right|
\leq \sum^N_{n=1} \int_\Omega|\bnabla\phi_n^{\omega}||\bnabla\bm\phi_n^{\sigma}|\,\mathrm{d}\bm x 
\leq \Theta\,.
\label{temp}
\end{equation}
The second term in \eqref{eq:trace3} satisfies
\begin{equation}
\Wi^{-1}\sum^N_{n=1} \int_\Omega |\bm\phi_n^{\sigma}|^2 \mathrm{d}\bm x = b\Wi^{-1}\,, \qquad 0<b\leq N\leq c\Theta^{1/2}\,.
\label{eq:traceWi}
\end{equation} 
The stretching term in \eqref{eq:trace4} that involves $\bnabla \bm v_n$ is first integrated by parts. The Cauchy-Schwarz inequality is then applied twice to obtain
\begin{equation}
\left\lvert\sum^N_{n=1} \int_\Omega \bm\phi_n^{\sigma}:[(\bnabla{\bm v}_n){\bm\sigma} +{\bm\sigma}(\bnabla{\bm v}_n)^{\top}]\,\mathrm{d}\bm x\right\rvert
\leq 2\sum^N_{n=1} \int_\Omega
(\lvert\bm\phi_n^\sigma\rvert\lvert\bnabla\bm\sigma\rvert\lvert\bm v_n\rvert+\lvert\bnabla\bm\phi_n^\sigma\rvert\lvert\bm\sigma\rvert\lvert\bm v_n\rvert)
\,\mathrm{d}\bm x.\nonumber
\end{equation}
We then use the same techniques employed to get \eqref{advectiveterms} and obtain\,:
\begin{equation}
\left|\sum^N_{n=1}\int_\Omega \bm\phi_n^{\sigma}:[(\bnabla{\bm v}_n){\bm\sigma} +{\bm\sigma}(\bnabla{\bm v}_n)^{\top}]\,
\mathrm{d}\bm x\right|
\leq c\,(1+\ln\Theta)^{1/2}\big(\Theta^{1/2}\|\bm\sigma\|_2+\Theta^{1/4}\|\bnabla\bm\sigma\|_2\big).
\label{eq:stretching_1}
\end{equation}
The other stretching term in \eqref{eq:trace4} is estimated by first applying the Cauchy--Schwarz inequality\,:
\begin{equation}
\left\lvert\sum^N_{n=1} \int_\Omega \bm\phi_n^{\sigma}:[(\bnabla{\bm u})\bm\phi_n^{\sigma}+\bm\phi_n^{\sigma}(\bnabla{\bm u})^\top]\,\mathrm{d}\bm x\right\rvert
\leq2\|\bnabla\bm u\|_2\left\lvert\int_\Omega\Big(\sum_{n=1}^N|\bm\Phi_n|^2\Big)^2\,\mathrm{d}\bm x\right\rvert^{1/2}.
\nonumber
\end{equation}
We note that in two dimensions $\|\bnabla\bm u\|_2=\|\omega\|_2$ and  then use the Lieb--Thirring inequality for the set of orthonormal functions $\{\bm\Phi_n\}_{n=1}^N$ \citep{CFT88} and \eqref{eq:traceoflaplacian} to obtain
\begin{equation}
\left|\sum^N_{n=1}\int_\Omega \bm\phi_n^{\sigma}:[(\bnabla{\bm u})\bm\phi_n^{\sigma}+\bm\phi_n^{\sigma}(\bnabla{\bm u})^\top]\,\mathrm{d}\bm x \right| \leq c\|\omega\|_2\Theta^{1/2}\,.
\label{eq:stretching_2}
\end{equation}
Using \eqref{laplacianterms}--\eqref{eq:stretching_2}, the trace $\operatorname{Tr}\left[{\mathsfbi{A}}{\mathsfbi{P}}_N\right]$ can now be estimated in terms of $\Theta$ as follows 
\begin{eqnarray}
\operatorname{Tr}\left[{\mathsfbi{A}}{\mathsfbi{P}}_N\right]
&\leq&\big(\beta\Wi^{-1}-1\big)\Rey^{-1}\Theta 
- b\Wi^{-1} 
+ c(1+\ln \Theta)^{1/2}\Theta^{1/2}\|\bm\sigma\|_2 \nonumber\\[1mm]
&&
+c(1+\ln \Theta)^{1/2}\Theta^{1/4}\left(\|\bnabla\bm\sigma\|_2+\|\bnabla\omega\|_2\right)
+ c\|\omega\|_2\Theta^{1/2}\,.
\label{eq:trace_2}
\end{eqnarray}
In the third term in the right-hand side of \eqref{eq:trace_2}, it is useful to write $\Theta^{1/2}=\Theta^{a}\Theta^{1/2-a}$ with $0 < a <1/2$. 
Taking the time average of \eqref{eq:trace_2} and applying the Cauchy--Schwarz and H\"older's inequalities on the time variable yields
\begin{eqnarray}
\langle\operatorname{Tr}\left[{\mathsfbi{A}}{\mathsfbi{P}}_N\right]\rangle
&\leq&\big(\beta\Wi^{-1}-1\big)\Rey^{-1}\langle\Theta\rangle
-b\Wi^{-1}
+c\langle\|\omega\|_2^2\rangle^{1/2}\langle\Theta\rangle^{1/2}\nonumber\\[1mm]
&&+c\langle(1+\ln \Theta)^{3/2}\Theta^{3a}\rangle^{1/3}\langle\Theta^{3(1/2-a)}\rangle^{1/3}\langle\|\bm\sigma\|_2^3\rangle^{1/3} 
\nonumber\\[1mm]
&& 
+c\langle(1+\ln \Theta)\Theta^{1/2}\rangle^{1/2}\big(\langle\|\bnabla\bm\sigma\|_2^2\rangle^{1/2}+\langle\|\bnabla\omega\|_2^2\rangle^{1/2}\big)\,.
\label{eq:timeave_1}
\end{eqnarray}
The bound in \eqref{eq:timeave_1} can be improved by estimating $\langle\|\bnabla\omega\|_2^2\rangle$. We first multiply \eqref{eq:NSE2} 
by $\omega$, integrate over space, and take the time average (noting that is necessary to use our assumption that $\|\omega\|_{2}$ is finite and hence $\langle\partial_t\|\omega\|_2^2\rangle=0)$\,:
\begin{equation}
\Rey^{-1}\langle\|\bnabla\omega\|_2^2\rangle
=\Big\langle\int_\Omega\omega\hat{\bm z}\bcdot\bnabla\times \bm F\,\mathrm{d}\bm x\Big\rangle
+ \beta(\Wi\,\Rey)^{-1}\Big\langle\int_\Omega\omega\hat{\bm z}\bcdot\bnabla\times(\bnabla\bcdot\bm\sigma)\,\mathrm{d}\bm x\Big\rangle\,.
\end{equation}
We then integrate by parts and apply the Cauchy--Schwarz inequality, first in space and then in time, to obtain
\begin{equation}
\langle\|\bnabla\omega\|_2^2\rangle^{1/2}
\leq \Rey\|\bm F\|_{2} + \beta\Wi^{-1}\langle\|\bnabla\bm\sigma\|_2^2\rangle^{1/2}.
\label{eq:gradomegabound}
\end{equation}
Moreover, Jensen's inequality allows us to express \eqref{eq:timeave_1} as a function of $\langle\Theta\rangle$\,:
\begin{eqnarray}\label{eq:timeave_2}
\langle\operatorname{Tr}\left[{\mathsfbi{A}}{\mathsfbi{P}}_N\right]\rangle
&\leq&\big(\beta\Wi^{-1}-1\big)\Rey^{-1}\mathcal \langle\Theta\rangle -b\Wi^{-1} 
+ c\langle\|\omega\|_2^2\rangle^{1/2} \langle\Theta\rangle^{1/2}\nonumber\\
&&+c(1+\ln\langle\Theta\rangle)^{1/2}\mathcal \langle\Theta\rangle^{1/2}\langle\|\bm\sigma\|_2^3\rangle^{1/3}
+c\Rey(1+\ln\langle\Theta\rangle)^{1/2}\mathcal \langle\Theta\rangle^{1/4}\|\bm F\|_2\nonumber\\
&&+c\big(1+\beta\Wi^{ -1}\big)(1+\ln\langle\Theta\rangle)^{1/2}\langle\Theta\rangle^{1/4}\langle\|\bnabla\bm\sigma\|_2^2\rangle^{1/2}.
\end{eqnarray}
Note that for this last inequality to hold, $a$ must be such that $\Theta^{3(1/2-a)}$ and $(1+\ln \Theta)^{3/2}\Theta^{3a}$ are concave functions.
The choice $a=1/5$ guarantees that $\Theta^{3(1/2-a)}$ is concave for all values of $\Theta$ and $(1+\ln \Theta)^{3/2}\Theta^{3a}$ is concave 
for $\Theta>5$. This restriction on $\Theta$ is justified since $\Theta\geq cN^{2}$ and in elastic turbulence $N$ is large. 
The trace $\langle\operatorname{Tr}\left[{\mathsfbi{A}}{\mathsfbi{P}}_N\right]\rangle$ is thus guaranteed to be negative if
$\langle\Theta\rangle$ satisfies
\begin{eqnarray}\label{eq:estimateN_1}
\left(1 - \beta\Wi^{-1}\right)\langle\Theta\rangle
&\geq& c\Rey\langle\|\omega\|_2^2\rangle^{1/2}\langle\Theta\rangle^{1/2} + 
c\Rey^2(1+\ln\langle\Theta\rangle)^{1/2}\langle\Theta\rangle^{1/4}\|\bm F\|_{2}\\
&& \hspace{-2cm}+c\Rey(1+\ln \mathcal \langle\Theta\rangle)^{1/2}\left\{\langle\Theta\rangle^{1/2}\langle\|\bm\sigma\|_{2}^{3}\rangle^{1/3} + \big(1+\beta\Wi^{-1}\big)
\langle\Theta\rangle^{1/4}\langle\|\bnabla\bm\sigma\|_{2}^{2}\rangle^{1/2}\right\}\,.\nonumber
\end{eqnarray}
Under the assumptions that $0<\Rey<\Rey_{\mathrm{c}}$ and $\Wi \gg 1$, it is easy to see that $(1 - \beta\Wi^{-1})\langle\Theta\rangle\approx\langle\Theta\rangle$ is greater than each of the terms in the first line of \eqref{eq:estimateN_1}. As $\|\bm F\|_2$ is independent of $\Wi$, $\langle\Theta\rangle$ dominates the forcing term. In the elastic-turbulence regime of the Oldroyd-B model, the kinetic-energy spectrum decays rapidly as a function of the wave number \citep{BBBCM08,BB10}, and $\|\omega\|_2$ is expected to be small as $\Wi$ increases. Therefore, $\langle\Theta\rangle$ also controls the enstrophy term in \eqref{eq:estimateN_1}\,---this is confirmed numerically in figure \ref{fig:comparison}.

The number of significant terms in \eqref{eq:estimateN_1} is thus reduced to two. Direct numerical simulations of the Oldroyd-B model show that 
in elastic turbulence $\operatorname{Tr}\bm\sigma$ grows as $\Wi$ increases and $\bm\sigma$ develops strong gradients \citep{BBBCM08,BB10}.
Therefore, both $\|\bm\sigma\|_2$ and $\|\bnabla\bm\sigma\|_{2}$ are expected to increase with $\Wi$. To analyze the $\bm\sigma$- and $\bnabla\bm\sigma$- terms in \eqref{eq:estimateN_1}, it is useful to note that for any $\mathcal{M}$ independent of $\langle\Theta\rangle^{1/2}\geq 4$ and any $0 \leq\lambda<3/2$ the following holds
\begin{equation}
\langle\Theta\rangle^{1/2}\geq c\mathcal{M}^{1/(2-\lambda)}(1+\ln\mathcal{M})^{1/2(2-\lambda)} \Longrightarrow
\langle\Theta\rangle\geq c\mathcal{M}\langle\Theta\rangle^{\lambda/2}(1+\ln\langle\Theta\rangle)^{1/2},
\label{eq:gettingN}
\end{equation}
which can be proved by generalizing an analogous result by \citet{DG91}.
The inequalities below are thus sufficient conditions for 
\eqref{eq:estimateN_1} to hold\,:
\begin{subeqnarray}
\langle\Theta\rangle^{1/2}&>& c\Rey\langle\|\bm\sigma\|_2^3\rangle^{1/3}
\big(1+\ln \Rey+\ln\langle\|\bm\sigma\|_2^3\rangle^{1/3}\big)^{1/2},
\slabel{eq:estsigmaapp}\\
%\mathcal{N} &\geq& c\Rey\langle\|\omega\|_2^2\rangle^{1/2}
%\slabel{eq:estomega}\\
%\mathcal{N} &\geq& c\Rey^{4/3}\|f\|_2^{2/3}
%\left(1+\ln \Rey^2+\ln\|f\|_2\right)^{1/3}\\
\langle\Theta\rangle^{1/2}&>& c
\Rey^{2/3}\langle\|\bnabla\bm\sigma\|_2^2\rangle^{1/3}
\big(1+\ln \Rey+\ln\langle\|\bnabla\bm\sigma\|_2^2\rangle^{1/2}\big)^{1/3}\,,
\slabel{eq:estgradsigmaapp}
\label{eq:estimationapp}
\end{subeqnarray}
Since $\Theta\geq cN^2$ (see \eqref{eq:traceoflaplacian}),
these can be converted into sufficient conditions on $N$,
as in \eqref{eq:estimation}.

%%%%%%%%%%%%%%%%%%%%%%
\bibliographystyle{jfm}
%\bibliography{tumblingjfmv01}

\begin{thebibliography}{14}
%\expandafter\ifx\csname natexlab\endcsname\relax\def\natexlab#1{#1}\fi
{

\bibitem[Amarouchene \& Kellay(2002)]{AK02}
{\sc Amarouchene, Y. \& Kellay, H.} 2002
Polymers in 2D turbulence: suppression of large scale fluctuations.
{\em Phys. Rev. Lett.}
{\bf 89}, 104502.
%doi:10.1103/PhysRevLett.89.104502
%link: http://journals.aps.org/prl/abstract/10.1103/PhysRevLett.89.104502

\bibitem[Barrett \& S\"uli(2011)]{BS2011}
{\sc Barrett, J.W. \& S\"uli, E.}
2011 
Existence and equilibration of global weak solutions to kinetic models 
for dilute polymers I. {\em Math. Models Methods Appl. Sci.} \textbf{21}, 1211--1289. 


\bibitem[Berti \etal(2008)]{BBBCM08}
{\sc Berti, S., Bistagnino, A., Boffetta, G., Celani, A. \& Musacchio, S.}
2008
Two-dimensional elastic turbulence.
{\em Phys. Rev.} E
{\bf 77}, 055306(R).
%doi:10.1103/PhysRevE.77.055306
%link: http://journals.aps.org/pre/abstract/10.1103/PhysRevE.77.055306

\bibitem[Berti \& Boffetta(2010)]{BB10}
{\sc Berti, S. \& Boffetta, G.}
2010
Elastic waves and transition to elastic turbulence in a two-dimensional viscoelastic Kolmogorov flow.
{\em Phys. Rev.} E
{\bf 82}, 036314.
%doi:10.1103/PhysRevE.82.036314
%link: http://journals.aps.org/pre/abstract/10.1103/PhysRevE.82.036314

\bibitem[Boffetta, Celani \& Musacchio(2003)Boffetta \etal]{BCM03}
{\sc Boffetta, G., Celani, A. \& Musacchio, S.}
2003
Two-dimensional turbulence of dilute polymer solutions.
{\em Phys. Rev. Lett.}
{\bf 91}, 034501.
%doi:10.1103/PhysRevLett.91.034501
%link: http://journals.aps.org/prl/abstract/10.1103/PhysRevLett.91.034501

\bibitem[Boffetta \etal(2005)]{BCMPV05}
{\sc Boffetta, G., Celani, A., Mazzino, A., Puliafito, A. \& Vergassola, M.}
2005
The viscoelastic Kolmogorov flow: eddy-viscosity and linear stability.
{\em J. Fluid Mech.}
{\bf 523}, 161--170.

\bibitem[Burghelea, Segre \& Steinberg(2004)Burghelea \etal]{BSS04}
{\sc Burghelea, T., Segre, E. \& Steinberg, V.}
2004
Statistics of particle pair separations in the elastic turbulent flow of a dilute polymer solution. 
{\em Europhys. Lett.}
{\bf 68}, 529--535.
%doi:10.1209/epl/i2004-10229-y 
%link: http://epljournal.edpsciences.org/articles/epl/abs/2004/22/epl8443/epl8443.html

\bibitem[Burghelea, Segre \& Steinberg(2007)Burghelea \etal]{BSS07}
{\sc Burghelea, T., Segre, E. \& Steinberg, V.}
2007
Elastic turbulence in von Karman swirling flow between two disks.
{\em Phys. Fluids}
{\bf 19}, 053104.
%doi:10.1063/1.2732234
%link: http://scitation.aip.org/content/aip/journal/pof2/19/5/10.1063/1.2732234

\bibitem[Constantin(1987)]{Const87} 
{\sc Constantin, P.} 
1987 
Collective $L^{\infty}$-estimates for families of functions with orthonormal derivatives. 
{\em Indiana Univ. Math. J.} \textbf{36}, 603--616.

%\bibitem[Constantin \etal(1985)]{CFMT85}
%\textsc{Constantin, P., Foias, C., Manley, O. P. \& Temam, R.} 1985
%Determining modes and fractal dimension of turbulent flows.
%\emph{J. Fluid Mech.} \textbf{150}, 427--440.
%doi:10.1017/S0022112085000209
%link=https://www.cambridge.org/core/journals/journal-of-fluid-mechanics/article/determining-modes-and-fractal-dimension-of-turbulent-flows/CA02F%3E52538A3897D49DFBCECA17CA1

\bibitem[Constantin \& Foias(1985)]{CF85}
{\sc Constantin, P. \& Foias, C.}
1985 Global Lyapunov exponents, Kaplan--Yorke formulas and the dimension of the attractors for 2D Navier--Stokes equations.
{\em Comm. Pure Appl. Math.}
{\bf 38}, 1--27.
%doi:10.1002/cpa.3160380102
%link: http://onlinelibrary.wiley.com/doi/10.1002/cpa.3160380102/abstract

\bibitem[Constantin, Foias \& Temam(1988)]{CFT88}
{\sc Constantin, P., Foias, C. \& Temam, R.} 1988 On the dimension of the attractors in two-dimensional turbulence. {\em Physica} D {\bf 30}, 
284--296.
%doi:10.1016/0167-2789(88)90022-X
%link: http://www.sciencedirect.com/science/article/pii/016727898890022X

%\bibitem[Constantin \& Sun(2012)]{CS12}
%{\sc Constantin, P. \& Sun, W.}
%2012
%Remarks on Oldroyd-B and related complex fluid models
%{\em Comm. Math. Sci.}
%{\bf 10}, 33--73.
%%doi: 10.4310/CMS.2012.v10.n1.a3
%%link: http://www.intlpress.com/site/pub/pages/journals/items/cms/content/vols/0010/0001/a003/

\bibitem[Constantin \& Kliegl(2012)]{CK12}
{\sc Constantin, P. \& Kliegl, M.}
2012
Note on global regularity for two-dimensional Oldroyd-B fluids with diffusive stress.
{\em Arch. Rational Mech. Anal.}
{\bf 206}, 725--740.
%doi: 10.1007/s00205-012-0537-0
%link: http://link.springer.com/article/10.1007/s00205-012-0537-0

%\bibitem[Constantin(2015)]{C15}
%{\sc Constantin, P.} 2015 Lagrangian--Eulerian methods for uniqueness in hydrodynamic systems. {\em Adv. Math.} {\bf 278}, 67--102.
%doi: 10.1016/j.aim.2015.03.010
%link: http://www.sciencedirect.com/science/article/pii/S0001870815001024

\bibitem[Doering \& Gibbon(1991)]{DG91}
{\sc Doering, C. R. \& Gibbon, J. D.}
1991
Note on the Constantin--Foias--Temam attractor dimension estimate for two-dimensional turbulence.
{\em Physica} D
{\bf 48}, 471--480.
% doi: 10.1016/0167-2789(91)90098-T
% link: http://www.sciencedirect.com/science/article/pii/016727899190098T

\bibitem[Doering \& Gibbon(1995)]{DG95}
\textsc{Doering, C. R. \& Gibbon, J. D.} 1995
\textit{Applied analysis of the Navier--Stokes equations}. Cambridge University Press.
%doi= http://dx.doi.org/10.1017/CBO9780511608803

\bibitem[Elgindi \& Rousset(2015)]{ER15}
{\sc Elgindi, T.~M. \& Rousset, F.}
2015
Global regularity for some Oldroyd-B type models.
{\em Comm. Pure Appl. Math.}
{\bf 68}, 2005--2021.
%doi: 10.1002/cpa.21563
%link: http://onlinelibrary.wiley.com/doi/10.1002/cpa.21563/abstract

\bibitem[El-Kareh \& Leal(1989)]{EL89}
{\sc El-Kareh, A.~W. \& Leal, L.~G.}
1989
Existence of solutions for all Deborah numbers for a non-Newtonian model modified to include diffusion.
{\em J. Non-Newtonian Fluid Mech.}
{\bf 33}, 257--287.
%doi: 10.1016/0377-0257(89)80002-3
%link: http://www.sciencedirect.com/science/article/pii/0377025789800023

\bibitem[Fouxon \& Lebedev(2003)]{FL03}
\textsc{Fouxon, A. \& Lebedev, V.} 2003
Spectra of turbulence in dilute polymer solutions.
\textit{Phys. Fluids} \textbf{15}, 2060--2072.
%doi:10.1063/1.1577563
%link:http://scitation.aip.org/content/aip/journal/pof2/15/7/10.1063/1.1577563

\bibitem[Gibbon \& Titi(1997)]{GT97}
{\sc Gibbon, J. D. \& Titi, E.}
1997
Attractor dimension and small length scale estimates for the three-dimensional Navier--Stokes equations.
{\em Nonlinearity}
{\bf 10}, 109--119.
%doi:10.1088/0951-7715/10/1/007
%link: http://iopscience.iop.org/article/10.1088/0951-7715/10/1/007

\bibitem[Grilli, V\'azquez-Quesada \& Ellero(2013)Grilli \etal]{GVQE13}
\textsc{Grilli, M., V\'azquez-Quesada, A. \& Ellero, M.} 2013
Transition to turbulence and mixing in a viscoelastic fluid flowing inside a 
channel with a periodic array of cylindrical obstacles.
\textit{Phys. Rev. Lett.} \textbf{110}, 174501.
%DOI: 10.1103/PhysRevLett.110.174501
%link: http://journals.aps.org/prl/abstract/10.1103/PhysRevLett.110.174501

\bibitem[Groisman \& Steinberg(2000)]{GS00}
{\sc Groisman, A. \& Steinberg, V.}
2000
Elastic turbulence in a polymer solution flow.
{\em Nature}
{\bf 405}, 53--55.
%doi:10.1038/35011019
%link: http://www.nature.com/nature/journal/v405/n6782/abs/405053a0.html

\bibitem[Groisman \& Steinberg(2001)]{GS01}
{\sc Groisman, A. \& Steinberg, V.}
2001
Efficient mixing at low Reynolds numbers using polymer additives.
{\em Nature}
{\bf 410}, 905--908.
%doi:10.1038/35073524
%link: http://www.nature.com/nature/journal/v410/n6831/full/410905a0.html

%\bibitem[Groisman \& Steinberg(2004)]{GS04}
%{\sc Groisman, A. \& Steinberg, V.}
%2004
%Elastic turbulence in curvilinear flows of polymer solutions.
%{\em New J. Phys.}
%{\bf 6}, 29.
%%DOI: 10.1088/1367-2630/6/1/029
%%link: http://iopscience.iop.org/article/10.1088/1367-2630/6/1/029/meta

\bibitem[Gupta, Perlekar \& Pandit(2015)Gupta \etal]{GPP15}
\textsc{Gupta, A., Perlekar, P. and Pandit, R.} 
2015
Two-dimensional homogeneous isotropic fluid turbulence with polymer additives.
{\em Phys. Rev.} E 
{\bf 91}, 033013.
%doi:10.1103/PhysRevE.91.033013
%link: http://journals.aps.org/pre/abstract/10.1103/PhysRevE.91.033013

\bibitem[Gupta \& Pandit(2017)]{GP16}
{\sc Gupta, A. \& Pandit, R.}
2017
Melting of a non-equilibrium vortex crystal in a fluid film with polymers: elastic versus fluid turbulence.
{\em Phys. Rev.} E
{\bf 95}, 033119.
%doi: arXiv:1602.08153
%link: https://arxiv.org/abs/1602.08153

\bibitem[Hu \& Lin(2016)]{HL16}
{\sc Hu, X. \& Lin, F.}
2016
Global solutions of two-dimensional incompressible viscoelastic flows with discontinuous initial data.
{\em Comm. Pure Appl. Math.}
{\bf 69}, 372--404.
%doi: 10.1002/cpa.21561
%link:http://onlinelibrary.wiley.com/doi/10.1002/cpa.21561/full

\bibitem[Kurganov \& Tadmor(2000)]{KT00}
\textsc{Kurganov, A. and Tadmor, E.}
2000
New high-resolution central schemes for nonlinear conservation laws and convection-diffusion equations.
{\em J. Comp. Phys.} 
{\bf 160}, 241--282.
%doi: ??????10.1006/jcph.2000.6459
%link: http://www.sciencedirect.com/science/article/pii/S0021999100964593

\bibitem[Larson(1992)]{L92}
{\sc Larson, R. G.} 1992
Instabilities in viscoelastic flows.
{\em Rheol. Acta} {\bf 31}, 213--263.
%doi:10.1007/BF00366504
%link:http://link.springer.com/article/10.1007/BF00366504

%\bibitem[Lieb \& Thirring(1976)]{LT76}
%{\sc Lieb, E. H. \& Thirring, W. E.}
%1976
%Inequalities for the moments of the eigenvalues of the Schr\"odinger Hamiltonian and their relation to Sobolev inequalities.
%In {\em Mathematical Physics}, pp. 269--303. Princeton University Press.
%%doi:
%%link=http://link.springer.com/chapter/10.1007%2F3-540-27056-6_16

\bibitem[Lei, Masmoudi \& Zhou(2010)]{LMZ10}
{\sc Lei, Z., Masmoudi, N. \& Zhou, Y.}
2010
Remarks on the blowup criteria for Oldroyd models.
{\em J. Differ. Equ.}
{\bf 248}, 328--341.
%DOI:10.1016/j.jde.2009.07.011
%link:http://www.sciencedirect.com/science/article/pii/S0022039609002708

%\bibitem[Lin \& Zhang(2005)]{LZ14}
%{\sc Lin, F.-H. \& Zhang, T.}
%2014
%Global small solutions to a complex fluid model in three dimensional.
%{\em Arch. Rat. Mech. Anal.}
%{\bf 216}, 905--920.
%%DOI:10.1007/s00205-014-0822-1
%%link:http://link.springer.com/article/10.1007/s00205-014-0822-1

\bibitem[Mitchell \etal(2016)]{MLHC16}
{\sc Mitchell, J., Lyons K., Howe, A. M. \& Clarke, A.}
2016
Viscoelastic polymer flows and 
elastic turbulence in three-dimensional porous structures.
{\em Soft Matter}
{\bf 12}, 460--468.
%DOI: 10.1039/C5SM01749A
%link: http://pubs.rsc.org/en/content/articlelanding/2015/sm/c5sm01749a#!divAbstract

\bibitem[Oldroyd(1950)]{O50}
{\sc Oldroyd, J.~G.}
1950
On the formulation of rheological equations of state.
{\em Proc. R. Soc.} A
{\bf 200}, 523--541.
%doi:10.1098/rspa.1950.0035
%link: http://rspa.royalsocietypublishing.org/content/200/1063/523

%\bibitem[Ray \& Vincenzi(2016)]{RV16}
%{\sc Ray, S. S. \& Vincenzi D.} 2016
%Elastic turbulence in a shell model of polymer solution.
%\emph{Europhys. Lett.} \textbf{114}, 44001.

\bibitem[Renardy (2006)]{R06}
\textsc{Renardy, M.} 2006
A comment on smoothness of viscoelastic stresses. 
\emph{J. Non-Newtonian Fluid Mech.} 
\textbf{138}, 204--205.

\bibitem[Robinson(2001)]{R01}
{\sc Robinson, J. C.} 2001 \textit{Infinite-dimensional dynamical systems.} Cambridge University Press.
%isbn: 9780521632041

%\bibitem[Robinson(2007)]{R07}
%{\sc Robinson, J. C.} 2007 Parametrization of global attractors, experimental observations, and turbulence. {\em J. Fluid Mech.} {\bf 578}, 495--507.
%DOI: 10.1017/S0022112007005137
%link: %https://www.cambridge.org/core/journals/journal-of-fluid-mechanics/article/parametrization-of-global-attractors-experimental-observations-and-turbu%lence/31CD733C4C47A5B44FD307E070072E4B

%\bibitem[Robinson(2013)]{R13}
%{\sc Robinson, J. C.} 2013
%Attractors and finite-dimensional behaviour in 2D Navier--Stokes Equations.
%In \emph{ISRN Mathematical Analysis} \textbf{2013}, 291823.
%%doi:10.1155/2013/291823

\bibitem[Shaqfeh(1996)]{S96}
{\sc Shaqfeh, E. S. G.}
1996
Purely elastic instabilities in viscometric flows.
{\em Ann. Rev. Fluid Mech.}
{\bf 28}, 129-185.
%DOI: 10.1146/annurev.fl.28.010196.001021
%link: http://www.annualreviews.org/doi/abs/10.1146/annurev.fl.28.010196.001021

\bibitem[Sureshkumar \& Beris(1995)]{SB95}
\textsc{Sureshkumar, R. \& Beris, A. N.} 1995
Effect of artificial stress diffusivity on the stability of numerical
calculations and the flow dynamics of time-dependent viscoelastic flows.
\textit{J. Non-Newtonian Fluid Mech.} \textbf{60}, 53--80.
%doi: 10.1016/0377-0257(95)01377-8

%\bibitem[Temam(1997)]{T97}
%\textsc{Temam, R.} 1997
%\textit{Infinite-dimensional dynamical systems in mechanics and physics.} Springer.
%%doi= http://dx.doi.org/10.1017/CBO9780511608803

\bibitem[Thomases \& Shelley(2007)]{TS07}
\textsc{Thomases, B. \& Shelley, M.} 2007
Emergence of singular structures in Oldroyd-B fluids. 
\emph{Phys. Fluids} 
\textbf{19}, 103103.

\bibitem[Thomases \& Shelley(2009)]{TS09}
\textsc{Thomases, B. \& Shelley, M.} 2009 
Transition to mixing and oscillations in a Stokesian viscoelastic flow. \emph{Phys. Rev. Lett.} 
\textbf{103}, 094501.
%doi:10.1103/PhysRevLett.103.094501

\bibitem[Thomases, Shelley \& Thiffeault(2011)Thomases \etal]{TST11}
\textsc{Thomases, B., Shelley, M. \& Thiffeault, J.-L.} 2011
A Stokesian viscoelastic flow: Transition to oscillations and mixing.
\textit{Physica} D \textbf{240}, 1602--1614.
%doi:10.1016/j.physd.2011.06.011

\bibitem[Thomases(2011)]{T11}
\textsc{Thomases, B.} 2011
An analysis of the effect of stress diffusion on the dynamics of creeping
viscoelastic flow.
\textit{J. Non-Newtonian Fluid Mech.} \textbf{166}, 1221--1228.
%doi:dx.doi.org/10.1016/j.jnnfm.2011.07.009

\bibitem[Vaithianathan \etal (2006)]{vaithi}
\textsc{Vaithianathan, T., Robert, A., Brasseur, J. G. \& Collins, L. R.} 2006
An improved algorithm for simulating three-dimensional, viscoelastic turbulence.
\emph{J. Non-Newtonian Fluid Mech.} \textbf{140}, 3--22.

%\bibitem[Watanabe \& (2014)]{WG14}
%\textsc{Watanabe, T. \& Gotoh, T.} 2014
%Power-law spectra formed by stretching polymers in decaying
%isotropic turbulence.
%\textit{Phys. Fluids} \textbf{26}, 035110.
}
\end{thebibliography}

\end{document}